# The morphology of cell spheroids in simple shear flow

**Rosalia Ferraro**[1,2], **Jasmin Di Franco**[1, **], **Sergio Caserta**[1,2,*], **Stefano Guido**[1,2]

[1]CEINGE Advanced Biotechnologies Franco Salvatore, Via G. Salvatore 436, 80131 Naples, Italy

[2]Department of Chemical, Materials and Production Engineering, University of Naples Federico II, P. le V. Tecchio 80, 80125 Naples, Italy

[*] **Correspondence:** Corresponding Author scaserta@unina.it

[**] **Present address**: Jasmin Di Franco, University of Vienna, Faculty of Physics, Boltzmanngasse 5, 1090, Wien

**Abstract**

Cell spheroids are a widely used model to investigate cell-cell and cell-matrix interactions in a 3D microenvironment *in vitro*. Most research on cell spheroids has been focused on their response to various stimuli under static conditions. Recently, the effect of flow on cell spheroids has been investigated in the context of tumor invasion in interstitial space. In particular, microfluidic perfusion of cell spheroids embedded in a collagen matrix has been shown to modulate cell-cell adhesion and to represent a possible mechanism promoting tumor invasion by interstitial flow. However, studies on the effects of well-defined flow fields on cell spheroids are lacking in the literature. Here, we apply simple shear flow to cell spheroids in a parallel plate apparatus while observing their morphology by optical microscopy. By using image analysis techniques, we show that cell spheroids rotate under flow as rigid particles. As time goes on, cells from the outer layer detach from the sheared cell spheroids and are carried away by the flow. Hence, the size of cell spheroids declines with time at a rate increasing with the external shear stress, which can be used to estimate cell-cell adhesion. The technique proposed in this work allows one to correlate flow-induced effects with microscopy imaging of cell spheroids in a well-established shear flow field, thus providing a method to obtain quantitative results which are relevant in the general field of mechanobiology.

## 1    Introduction

Cell spheroids are a widely used model to investigate cell-cell and cell-matrix interactions in a 3D microenvironment *in vitro*. Due to their structural and functional similarities with *in vivo* tumors [1-3], spheroids have emerged as a valuable tool for cancer research and drug testing *in vitro*. One of the main advantages of cell spheroids is that they can be prepared by well-established culture protocols, such as the hanging drop method, micro-molding and rotary cell culture systems, which exploit cell aggregation in the absence of adhesive surfaces [4-8]. The initial cell aggregates grow as a sigmoidal function of time, with an initial exponential phase, an intermediate linear growth, and a tendency to reach a plateau at later times [9]. While growing, the cell spheroids undergo a change in their internal microstructure, since at some point diffusion limitation of nutrients, such as oxygen and glucose, and of metabolic waste tends to inhibit cell proliferation inside the spheroid. Therefore, an inner necrotic core develops in the central region of cell spheroids in addition to an outer proliferating and an intermediate non-proliferating layer [10-12]. This complex, heterogeneous microstructure poses several challenges to the study of cell spheroids in response to external factors, such as chemical and mechanical stimuli. The deformability of these structures [13] has been investigated through various techniques, including isotropic compression by osmotic pressure [14,15], atomic force microscopy (AFM) [16,17], and micro-indentation. The experimental results have been interpreted in terms of an equivalent surface tension, reflecting the attractive interactions between cells which hold the spheroid together.

While most of the research on cell spheroids has been focused on static conditions, the effect of flow has recently been investigated in the context of tumor invasion in interstitial space. In particular, microfluidic perfusion of cell spheroids embedded in a collagen matrix has been shown to modulate cell-cell adhesion and to represent a possible mechanism promoting tumor invasion by interstitial flow [18]. However, studies on the effects of well-defined flow fields on cell spheroids, which can provide a quantitative characterization of their response to mechanical stimuli, are lacking in the literature.

Here, we apply a simple shear flow to cell spheroids in a parallel plate apparatus while observing their morphology by optical microscopy. By using image analysis techniques, in this Brief Research Report we provide the first characterization of the flow behavior of cell spheroids in a simple shear velocity field.

## 2    Materials and Methods

The cell spheroids are obtained from NIH/3T3 (ATCC Cat# CRL-1658, RRID: CVCL_0594) mouse fibroblasts cultured in Dulbecco's Modified Eagle's Medium (DMEM) supplemented with 10% (v/v) Fetal Bovine Serum (FBS), 1% (v/v) antibiotics (50 units/mL penicillin and 50 mg mL-1 streptomycin) and 1% (v/v) L-glutamine. Cells harvested from monolayer cultures were seeded in a multiwall plate pre-coated with non-adhesive agarose at a concentration of $8 \cdot 10^3$ cells/well and covered with the cell growth medium. The multiwell plate was then incubated under typical cell culture conditions at 37 °C for 5 days to obtain compact spheroids. The culture protocol is schematically illustrated in **Figure 1A**.

The so obtained cell spheroids were then transferred in an aqueous solution of polyvinylpyrrolidone (PVP, from Sigma Aldrich, with number average molecular weight Mn 360kDa, CAS Number 9003-39-8), with concentrations in the range 20-27% wt. The addition of PVP enables to change the viscosity η of the cell spheroids suspending medium and thus to modulate the applied shear stress τ = ηγ̇, where γ̇ is the shear rate. Furthermore, higher viscosities of the suspending medium slow down sedimentation of cell spheroids. The viscosity of the PVP solutions was measured as a function of the shear rate in a rotational stress-controlled rheometer (Anton Paar Physica MCR 301 Instruments) equipped with a plate-cone (CP75-1/TI/S-SN18730). The rheological tests were run at room temperature (23°C). The plot of viscosity *vs* shear rate for the PVP solutions used in this work is shown in **Figure 1B** and is characterized by a Newtonian plateau at low shear rates (up to about 10 s$^{-1}$ for all the samples, with values of η from 10 to 60 Pa·s) followed by a shear thinning region. In the range of shear rates of our experiments, 0.5 – 5 s$^{-1}$, elastic effects can be considered as negligible and the rheological behavior of the PVP solutions is essentially Newtonian.

The PVP solutions with cell spheroids were loaded in a parallel plate apparatus which has been described in detail elsewhere [19] and is schematically shown in **Figure 2A**.

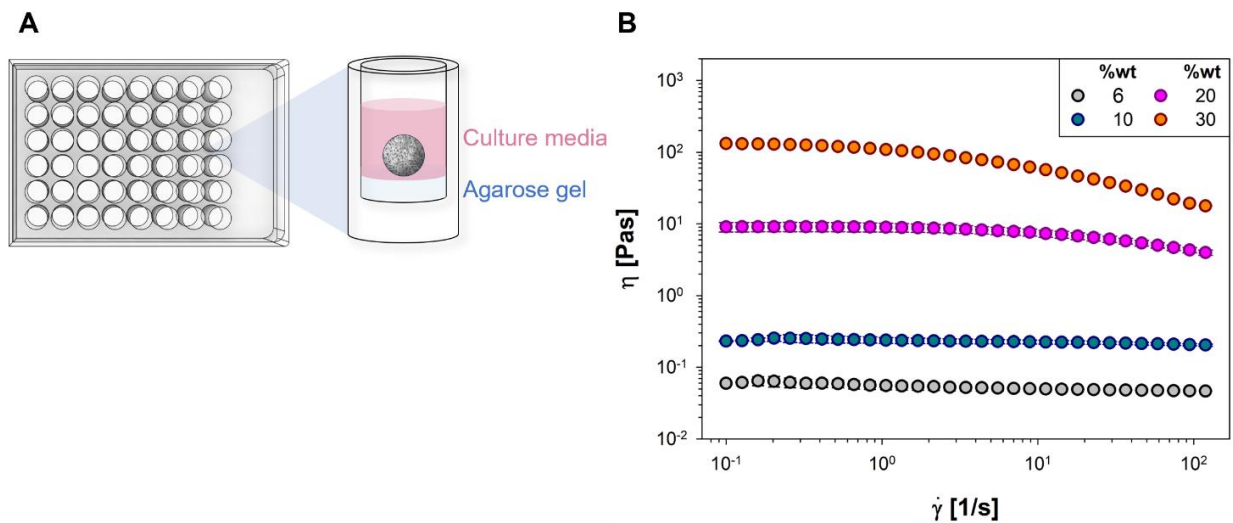

**Figure 1. A)** Illustration of the culturing method for NIH/3T3 mouse fibroblast spheroids using multiwell plates, with spheroids embedded in agarose gel within a culture media environment; **B)** Viscosity, η, *vs* shear rate, $\dot{\gamma}$, of PVP solutions at different weight concentrations (6, 10, 20 and 30%wt, in grey, blue, pink, and orange symbols, respectively).

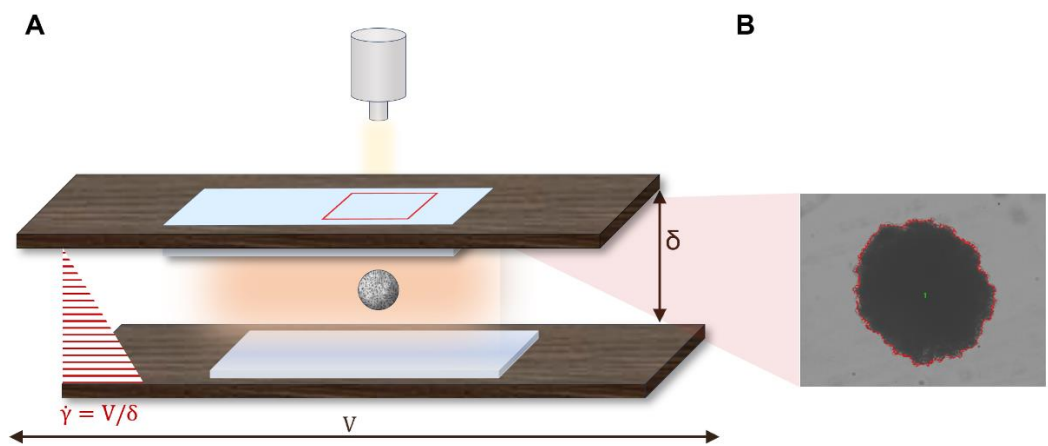

**Figure 2**. **A)** Schematics of the parallel plates used to apply simple shear flow to cell spheroids. The shear rate, $\dot{\gamma}$, is defined by the relative velocity, V, between plates and their separation distance, δ; **B)** Typical image of a cell spheroid in bright field microscopy with a red outline of the boundary as obtained from threshold segmentation.

Briefly, the apparatus is made of two rectangular glass plates inserted in windows cut on two rigid mounts. Parallelism between the two plates is adjusted by a set of micrometric rotary and tilting stages (the residual error is around 20 μm over a length of 10 cm). Shear flow is obtained by translating one of the plates (specifically, the bottom plate) with respect to the other by a computer-controlled motorized translating stage with micrometric precision. The microscope itself is mounted on a motorized translating stage that is used to keep the flowing spheroid within the field of view during motion. The microscope is equipped with a video camera (Hamamatsu, ORCA-spark Digital CMOS camera C11440-36U) for image acquisition during flow. The shear rate $\dot{\gamma}$ is equal to the slope of the linear velocity profile shown in **Figure 2A** and can be calculated as V/δ, where V is the speed of the moving plate and δ is the gap between the plates. The latter is set by a vertical translating micrometric stage and checked by focusing the surfaces of the glass plates. Typical values of δ in our experiments are in the range 300-500 μm. Quantitative analysis of spheroid shape is performed by using standard image analysis routines based on threshold segmentation to identify the spheroid contour from the surrounding background. A typical example of the image analysis outcome is presented in **Figure 2B**, where the red line is the spheroid contour. The geometrical features measured by image analysis include the area and the axes of the cell spheroid.

## 3  Results and discussion

In the range of shear stress investigated, σ ∈ [20-670] Pa – selected based on published outcomes [20] and to ensure the osmotic pressure of PVP on spheroids is negligible – the cell spheroids exhibited minimal, if any, deformation under shear flow [21]. Instead, they underwent three-dimensional rotational motion which is illustrated in the sequence of images in **Figure 3A**, which refers to a shear rate of about 1 s$^{-1}$ and a cell spheroid with a radius of 170 μm.

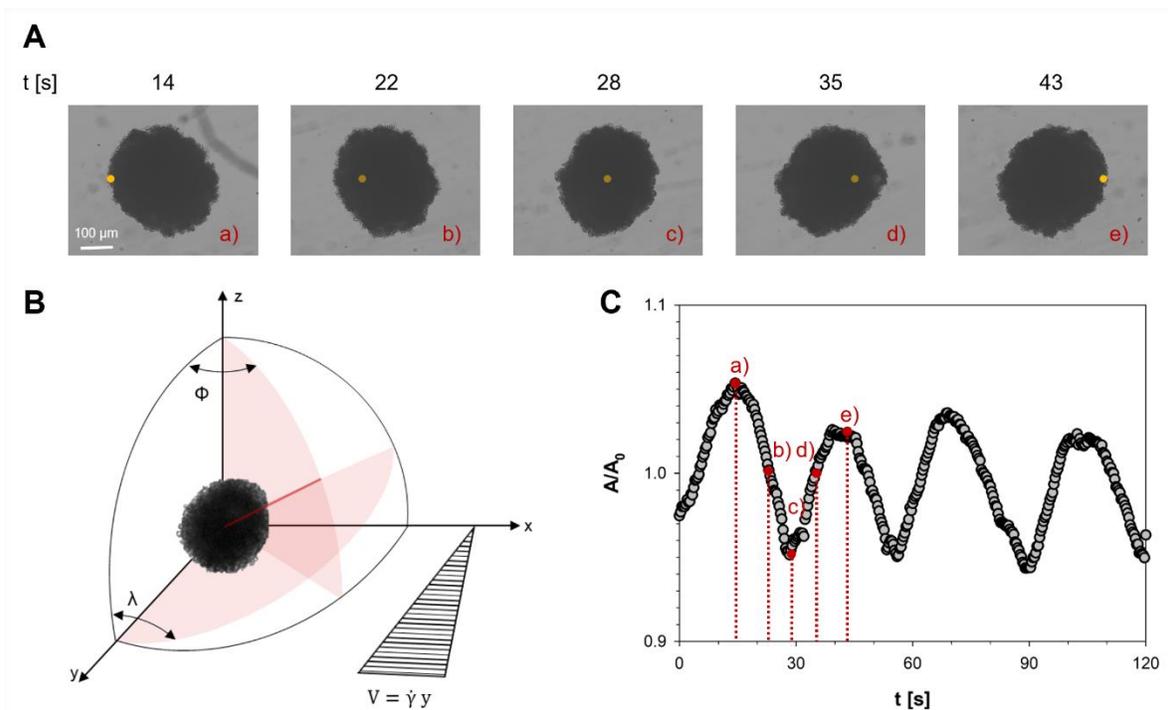

**Figure 3. A)** Time-lapse images of a cellular spheroid subjected to shear flow at indicated time points. The yellow points highlight the spheroid's rotation; **B)** The coordinate system employed to characterize the spheroid's rotational motion under a shear flow **V = γ̇ y**; **C)** Normalized projected area data of the cellular spheroid, A/A$_0$, plotted as a function of time, t. Time points corresponding to the images in panel **A** are marked on the graph (a-e) with red dashed lines.

The sequence is part of a video which is available as Supplementary Material (**Video S1**). The video has been made by assembling single images taken under flow and centering the spheroid in the field of view (Lagrangian reference). The images in **Figure 3A** apparently show that the cell spheroid rotates in the plane of observation, which is perpendicular to the velocity gradient. As a matter of fact, the video shows that the spheroid is actually rotating in the shear plane (x-y in **Figure 3B**) and that the apparent rotation of the spheroid axis in the plane of observation (yellow points in **Figure 3A**) is due to the irregular, rough shape of the surface, which exhibits several bumps and valleys. The combined effect of the spheroid rotation in the x-y plane and the coming in and out of focus of the surface roughness generates the apparent change of the projected area during flow (**Figure 3C**). The observed rotation is characteristic of rigid particles in a shear flow field, which have been theoretically studied for ellipsoidal particles by Jeffery under the assumptions of no sedimentation, negligible inertia, rigid particles and no slip at the solid-liquid interface [22] and are schematically represented in **Figure 3B**. In the latter, $\phi$ is the angle of rotation around the polar axis z, which is perpendicular to the shear plane x-y. The motion of the cell spheroid in **Figure 3A** is the projection of the long axis in the x-z plane of observation, which describes the angle $\lambda$ (equal to 0° in our case) in the same plane. The predictions of Jeffery have been confirmed later on by experimental results on cylindrical particles in a paper by Trevelyan and Mason [23], from which **Figure 3B** has been adapted. Further evidence of the rotary motion of the cell spheroids comes from the analysis of the time evolution of its shape. In **Figure 3C** the projected area of the cell spheroid A, normalized with respect to the initial area $A_0$, is plotted as a function of time for the cell spheroid of **Figure 3A**. The data of the normalized area fluctuates around an average value with a 4% amplitude oscillation and a period T of approximately 60 s. Such value of the period is larger than the one calculated from the expression derived by Jeffery in the case of a spherical particles $T = \frac{4\pi}{\dot{\gamma}}$, which is equal to 12.9 s for the cell spheroid in **Figure 3A**. The values calculated from the equation of the period for a prolate ellipsoid $T = \frac{2\pi(a^2+b^2)}{ab\dot{\gamma}} = \frac{2\pi(1+r^2)}{r\dot{\gamma}}$, where a and b are the major and minor axis of the ellipsoid, respectively, and r = a/b is the aspect ratio, are equal within experimental error to the ones for a spherical particle given the almost spherical shape of the cell spheroids. For example, the cell spheroid in **Figure 3A** has an aspect ratio r = 1.05, which corresponds to T = 12.83 s. It has been shown (see experimental results on ferric hydroxide flocs [24] and theoretical predictions for porous particles under shear flow [25]) that the agreement between theory and experiments is not affected by the fact that particles have a rough surface (see the images in **Figure 2B** and **Figure 3A**) as opposed to the smooth one considered in Jeffery's analysis. Therefore, the discrepancy between the value of the period measured experimentally and the one calculated from Jeffery's theory is likely due to the strong confinement of the spheroid between the parallel plates, since the gap is close to the diameter of the spheroid.

The $A/A_0$ data in **Figure 3C** seem to exhibit a declining trend in the 3 minutes time frame of the diagram. Results from a time frame of 25 minutes from the same experiment as **Figure 3C** are plotted in **Figure 4A**, where the decreasing trend of $A/A_0$ is more evident. A similar trend has been found for all the cell spheroids investigated in this work.

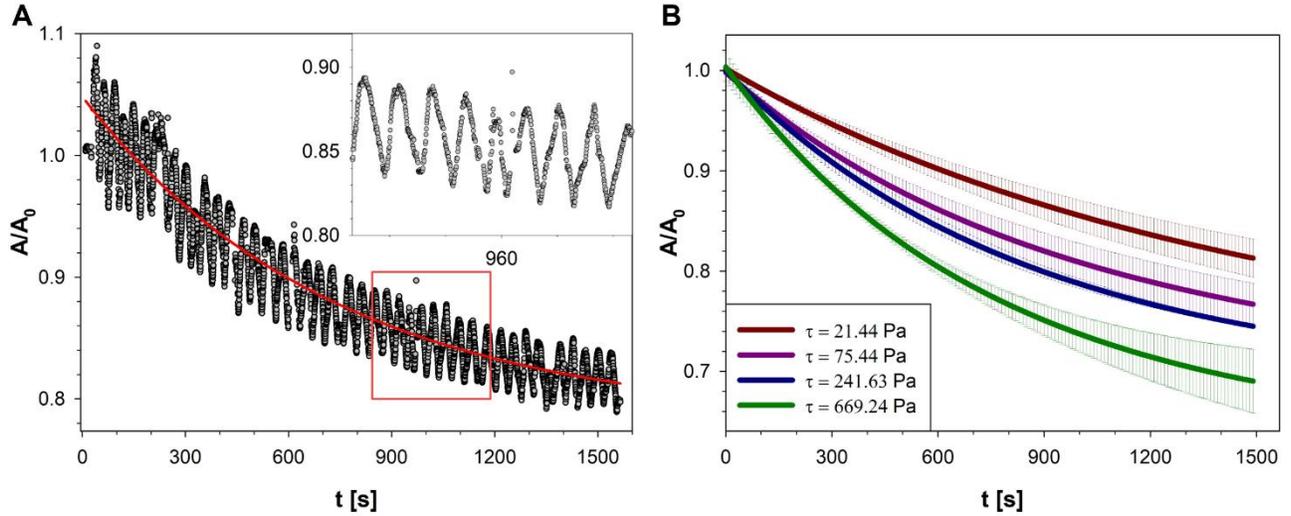

**Figure 4. A)** The normalized projected area of the cell spheroid of **Figure 3A** as a function of time, at $\dot{\gamma}$=1 1/s and τ=29 Pa; a zoomed view of the circled region is given in the inset, and the continuous line represents the average of the data; **B)** The time evolution of the normalized projected area of cell spheroids under different values of the external shear stress.

A careful examination of the images taken during shear flow shows indeed that cells from the outer layer tend to detach from the cell spheroid and are taken away by the suspending medium. This phenomenon presents some analogy with the dissociation of cells from spheroids undergoing perfusion, which has been attributed to flow-induced downregulation of cell-cell adhesion by reduced expression and altered localization of E-cadherin [26]. It can be speculated that a similar mechanism is at play in our experiments and that cells detachment from the spheroid is due to downregulation of cell-cell adhesion, although more work is needed to check this hypothesis. While losing cells from the edge and reducing their size, the cell spheroids keep rotating as a rigid particle with almost the same period, as shown by the oscillations in the inset of **Figure 4A**. To facilitate interpretation, raw data from each spheroid were smoothed to yield an averaged curve, represented by the continuous red line in **Figure 4A**.

The effect of the applied shear stress was studied by changing the shear rate and/or the viscosity of the suspending solutions. The outcomes are presented in **Figure 4B**, where the normalized area is plotted as a function of the shear stress. For the sake of clarity, only the average curves are shown in **Figure 4B** as red lines with the gray regions representing the standard error of mean evaluated at least on 5 spheroids. Under constant stress, the normalized area reduces over time until it stabilizes at a 'plateau'—a phase where the strain does not significantly fluctuate (with an error margin of ≤ 5% relative to a preceding measurement). At any measured interval, the data consistently shows a decline in $A/A_0$ correlating with an increase in shear stress, ranging from 21 to 699 Pa. This finding is in line with the interpretation of the normalized area reduction with time as a result of flow-induced erosion of the surface.

## 4    Conclusions

In this Brief Research Report, we show for the first time the flow behaviour of cell spheroids in a viscous fluid under simple shear flow. The experiments were run in a parallel plate apparatus while observing their morphology by optical microscopy. The results provide strong evidence that cell spheroids rotate as rigid particles under the action of shear flow, with a period in close agreement with the predictions of Jeffery's theory of ellipsoidal particles. A similar flow behavior has been observed for vesicles and red blood cells under shear flow [27,28]. In addition, a progressive detachment of cells from the outer spheroids layer was observed in all the experiments. As a consequence, the size of the cell spheroids decreased as a function of time in a shear stress-dependent fashion, due to detachment of cells from spheroids. No apparent change of T as a consequence of spheroid erosion was found, likely due to the fact that the period is rather insensitive to the aspect ratio in the range investigated in this study. The good agreement with the theoretical values of T was rather robust also taking into account other effects, which were outside the assumptions of Jeffery's theory. One of such effects is the irregular, rough surface of the cell spheroids, which did not affect the rotation period, in line with the previous results from the literature on particle flocs. Another effect is the influence of the confining walls since the ratio between cell spheroid diameter and gap size was around 0.8. However, the agreement of the experimental values of T with the ones predicted for unbounded flow suggest that wall effects are rather limited. This finding is in agreement with previous results from the literature on rods under shear flow, where little effect of wall proximity on the rotation period was found both in experiments [29] and numerical simulations [30]. The results of our work are relevant in the general field of mechanotransduction, which is addressed to study the mechanisms used by cells to convert mechanical stimuli in a biological response. As an example, future directions of this study include a biological analysis of the cell response under the applied shear stress to investigate the molecular drivers of cell detachment and to assess an equivalent surface tension [13,31] that could serve as a biomarker for differentiating the invasiveness of various cell lines, thereby paving the way for personalized medical strategies. Currently, our observations of cell detachment from spheroids under stress reveal a complex interplay that transcends mere deformation, challenging the simplistic extrapolation of the liquid drops model for surface tension. Acknowledging this intricacy, we are committed to further exploring the mechanical behavior after detachment. Our forthcoming investigations will particularly focus on whether spheroids experience deformation under increased stress levels. Such studies will augment our understanding of surface tension and its broader biological implications, reinforcing the foundational goals of our research. Furthermore, the analysis of the shear flow of cell spheroids in the context of Jeffery's theory is also relevant for the application of the latter theory to active biological matter, such as micro swimmers, which has recently experienced a revival of interest [32]. The study of the motion of cell spheroids in shear flow can also be extended by using different cell lines and by exploring wider ranges of shear rate, but all these activities are outside the scope of this Brief Research Report and can be the subject of future work.

## 5 Conflict of Interest

The authors declare that the research was conducted in the absence of any commercial or financial relationships that could be construed as a potential conflict of interest.

## 6 Author Contributions

Conceptualization: SC, SG; Data curation: RF; Formal analysis: RF, JDF; Investigation: RF, SC, SG; Methodology: SC, SG; Supervision: SC; Validation: RF; Visualization: RF, SG; Writing - original draft preparation: RF; Writing - original draft preparation, review and editing: SG.

## 7 Funding

This research received no specific grant from any funding agency in the public, commercial, or not-for-profit sectors.

This research received no specific grant from any funding agency in the public, commercial, or not-for-profit sectors.

## 8	Data Availability Statement

The raw data supporting the conclusions of this article will be made available by the authors, without undue reservation.